# Full field-of-view multi-targets imaging through scattering beyond 3D optical memory effect

WEI LI, JIETAO LIU, SHUNFU HE, LIXIAN LIU AND XIAOPENG SHAO*

*School of Physics and Optoelectronic Engineering, Xidian University, Xi'an, 710071, China*

**A robust method and strategy for efficient full field-of-view and depth separation optical imaging through scattering media regardless of the three-dimensional (3D) optical memory effect are proposed. In this method, the problem of imaging de-aliasing, decomposition, and separation of speckle patterns are solved taking advantages of the spatial decorrelation characteristics of speckles by employing randomly modulated illumination strategy and independent component analysis methods. Full field-of-view imaging of multi-targets locate at diverse spatial positions behind a scattering layer are realized and observed experimentally, for the first time, to the best of our knowledge. The method and strategy provide a potentially useful means for incoherent imaging through scattering in a wide class of fields such as optical microscopy, biomedical imaging, and astronomical imaging.**

Optical memory effect (OME) [1-3] dealing with scattering has been developed into a very significant and practical effect for imaging through or inside disordered media [4, 5], and for controlling the spatial and spectral shape of the transmitted light [6-8]. However, the severely constrained field-of-view (FOV) and difficulty in treating strong scattering situations hinder the development of the technology for further applications. Different approaches have been proposed aiming to expand the FOV for scattering imaging. The physical strategy can be roughly three-pronged. First, to characterize the medium piecemeal as well as possible by utilizing diverse pre-calibrated point spread function (PSF) laterally and longitudinally so that the large FOV is gained by stitching multi-shot and multi-view images [9, 10]. Second, to exploit additional prior knowledge whether it be straightforward or obscured, with that the entangled signals can be stripped away from raw captured images [11, 12]. Third, to build a complicated nonlinear relationship between input and output field assisted by eigenchannels or higher-order interconnections [13-15]. The pre-calibrated method is time-consuming and impractical. The prior-information approach is useful, however, the aliasing or mixing of speckles is unresolvable. The third method is showing promising ways for manipulating light for imaging, however, the measurement of transmission matrix requires that the objects and the scattering samples remain completely stationary over the long acquisition sequence. The deep-learning methods are also showing exciting results recently for optical imaging through turbid medium [16, 17], however, a huge number of input-output training pairs are required. The method is accompanied by problems of the insufficient generalization ability of the network, and the disability at solving multi-objects especially in the reconstruction of complex objects. How to break the limit of the OME, and how to enlarge the FOV for imaging multi-targets through scattering are significant issues to be addressed, which are highly desired by the researchers for not only the physic but also the practical applications.

Within a single OME zone, the imaging process can be simplified into a linear convolution model [18]. However, objects beyond OME region will inevitably produce speckles overlap. Most existing imaging techniques mentioned above regard OME as a limit and aim at coping with the unavoidable decline in correlation inherently introduced by the turbid medium.

Here, in contrast, we utilize the decorrelation characteristics of speckle patterns generated from different positions and objects beyond the OME region. Based on the independence of the different speckles' spatial information, the isolated speckles forming the mixed patterns can be extracted and retrieved via independent component analysis (ICA) [19, 20]. Full FOV imaging and depth separation of multi-objects at different spatial locations regardless of 3D OME behind scattering medium are achieved and experimentally observed. Moreover, we explore the de-aliasing properties of the proposed method mathematically indicated by the correlation coefficients and scatter diagrams. These properties give important insight into optical properties of the diffuser and clarify the required operating conditions of the proposed method.

Assuming an object $O(\vec{r})$ consists of $n$ discrete targets $O(\vec{r}_i)$ situated at different spatial positions, and the corresponding PSFs for each target are $H(\vec{r}_i)$. For incoherent illumination and due to the cross-correlation feature between different PSFs through scattering medium [14], the scrambled speckle reads [21]:

$$I = \sum_{i=1}^{n} O(\vec{r}_i) * H(\vec{r}_i), \qquad (1)$$

where ' $*$ ' denotes convolution operation. Inevitably, chaotic speckle signals generated from diverse objects' positions overlapped with each other. Directly solving this underdetermined equation seems impossible.

To accomplish this and avoid multi-view detectors signal acquisition scheme, we employ a strategy (see Fig. 1) by randomly adjusting the illumination ratio on each target, for example and for simplicity, via a beam attenuator, which is designed to work as a linear mixture modulator:

$$I_k = \sum_{i=1}^{n}\left[O(\vec{r}_i)\cdot S_k(\vec{r}_i)\right]*H(\vec{r}_i), \quad (2)$$

where $S_k(\vec{r}_i)$ denotes the illuminate part and subscript $k$ denotes the $k$th measurement. Since the attenuation plate can only modulate the input light intensity, the modulating illumination can be equivalent to a weighted constant:

$$I_k \approx \sum_{i=1}^{n} a_{ki} O(\vec{r}_i) * H(\vec{r}_i). \quad (3)$$

By setting $T_j = O(\vec{r}_j) * H(\vec{r}_j)$ and choosing $n$ equals to $k$, the matrix notation derived from (3) can be expressed as:

$$\begin{bmatrix} I_1 \\ I_2 \\ \vdots \\ I_n \end{bmatrix} = \begin{bmatrix} a_{11} & a_{12} & \cdots & a_{1n} \\ a_{21} & a_{22} & \cdots & a_{2n} \\ \vdots & \vdots & \ddots & \vdots \\ a_{n1} & a_{n2} & \cdots & a_{nn} \end{bmatrix} \begin{bmatrix} T_1 \\ T_2 \\ \vdots \\ T_n \end{bmatrix}. \quad (4)$$

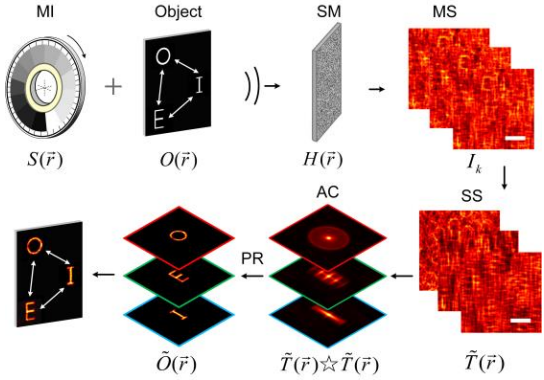

Fig. 1. Pipeline for multi-targets imaging through scattering regardless of OME. MI: modulate illumination; SM: scattering medium, MS: mixed speckles; SS: separated speckles; AC: autocorrelation; PR: phase retrieval.

Mathematically, the formula (4) can be simplified as:

$$\boldsymbol{I} = \boldsymbol{A}\boldsymbol{T}. \quad (5)$$

Thanks to the incoherence feature of $H(\vec{r}_i)$ and by exploiting the ICA based di-mixing model (see detail in appendix), well-separated speckle components can be formulated as:

$$\tilde{\boldsymbol{T}} = \boldsymbol{W}\boldsymbol{I} = \boldsymbol{W}\boldsymbol{A}\boldsymbol{T}. \quad (6)$$

$\boldsymbol{W}$ is a $n \times n$ separating matrix generated by maximizing the non-Gaussianity of the probability density function of signals [20]. The superscript ' $\sim$ ' means an estimated edition of the variable.

Finally, we obtain image of the sub-objects $\tilde{O}(\vec{r}_j)$ by calculating the autocorrelation of each isolated speckle component independently $\tilde{T}(\vec{r}_j) \star \tilde{T}(\vec{r}_j)$, and then the recovery of objects can be obtained by employing iterative phase retrieval (PR) algorithm [22]. Each part of the scenario with every single target can be recovered efficiently and effectively.

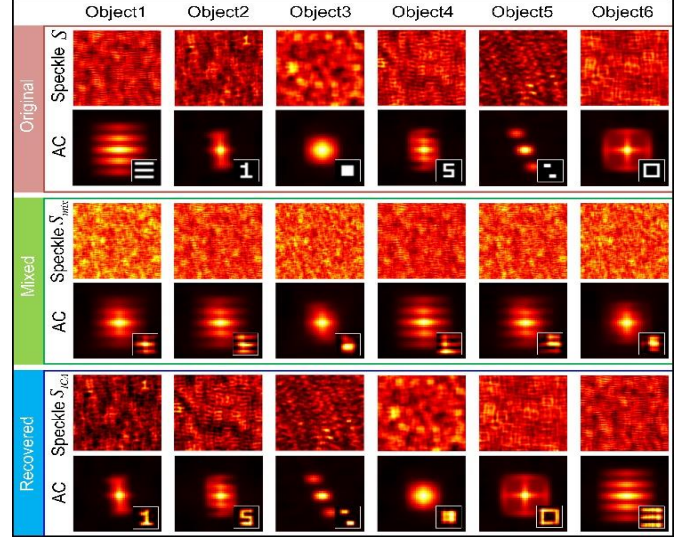

Fig. 2. Numerical investigations of multi-targets imaging beyond OME directly using PR and using the proposed method.

To numerically investigate the operation of the blind speckle separation, we performed simulations (Eq. (3)) using binary target samples (insets in the second row of Fig. 2). Objects located at different positions form the mixed speckle pattern $S_{mix}$ with different ratios are shown in the third row of Fig. 2, where the object's autocorrelation cannot be preserved, and imaging recovery directly using PR algorithm fails (see insets in the fourth row of Fig. 2). Using the proposed de-mixing strategy, speckles $S_{ICA}$ from different objects can be clearly separated, and successful imaging with excellent performance is realized (inset in the last row in Fig. 2).

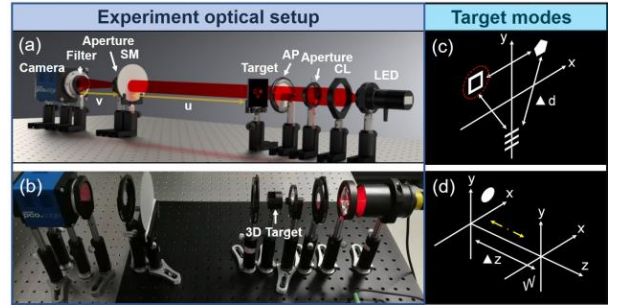

Fig. 3. Schematic of experimental setup and imaging exceeding lateral or longitudinal OME range. CL: collimating lens; AP: attenuation piece; SM: scattering medium.

In order to verify the proposed method, group of experiments are carried out. The schematic and the real experimental setup are shown in Fig. 3(a) and (b). A tailored object, fabricated by plastic film printing as a set of targets, is back-illuminated by a red-light LED (M625L3, Thorlabs). Light coming from the objects passes through a diffuser (220-grit ground glass, Edmund), an iris and a narrowband filter (633FS02-50, Andover), then forms a scrambled pattern on the detector (PCO. edge 4.2, pixel size $p$: 6.5μm). Targets mounted laterally and axially with centimeter-scale separations are shown in Fig. 3(c-d). The distance $u$ between the object and the scattering sample is set as 24 cm in (c) mode, while the distances for three axially separated targets are 22, 23, 24cm, respectively. The image distance $v$ in all experiments is 12cm.

Relative to conventional snapshot speckle correlation (SC) imaging setups [11], to obtain modulated illumination incidence, the attenuation plate (GCC-303002, DaHeng Optics) is used. By randomly rotating the attenuation piece while avoiding repetitions, one can easily get a series of instantaneous liner mixed speckles in different weighted factors.

The speckle pattern $i$ for the $i$th input mode can be written as 1× $10^6$ vectors. These vectors then constitute the $i$th row of $I$, which is a $k \times 10^6$ matrix, as described in formula (4). The derivation of the ICA based algorithms requires that the statistical feature of data is centered and pre-whitened, resulting in new vectors $z$. Based on the central limit theorem, the problem now is changed to find a de-mixing matrix $W$ on the speckle $I$ to maximum the non-Gaussianity of the de-mixed speckle $\widetilde{T}$, as described in formula (6). More precisely, the ICA of a vector $i$ searches for a linear transformation which minimizes the statistical dependence between its components $t$ (row vectors. Matrices are noted with capital bold letters $T$). The steps of the algorithm are given in appendix A.

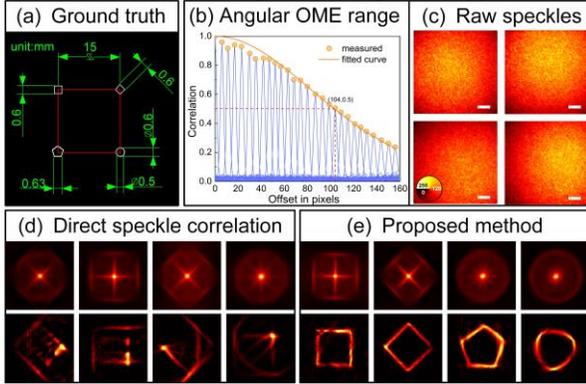

Fig. 4. Experiment results of imaging beyond angular OME through scattering layer. Scale bars are 2mm.

To verify the ability of the proposed method for imaging beyond OME, we quantify the effective FOV in experiment (Fig. 3(a)) by measuring the angular OME range (Fig. 4(a)). A 100$\mu m$ pinhole is used as a point source placed at a distance $u$=24cm away from the sample and scanned perpendicular to the optical axis direction by a translation stage (MTS50A-Z8, Thorlabs) with a fixed 0.1mm step. Speckles are captured at a distance $v$=12cm behind the sample. The effective OME range is evaluated by:

$$\Delta X_{FOV} = 2\beta p \Delta x, \qquad (7)$$

where $p$ denotes detector's pixel size, $\beta = u/v$ is the magnification of the imaging system, and $\Delta x$ implies the offset pixels in the image. We use the correlation function in [4] for curve fitting.

As shown in Fig. 4(b), the half width at half maximum (HWHM) of the fitted line for speckles-correlation reads 104 pixels. The calculated value of lateral OME range is 2.7 mm. As depicted in Fig. 4(a), the distance among targets used in this experiment is 15mm, which is 5.56 times of the lateral OME range. The full-FOV imaging can be achieved regardless of the OME range providing that the overlapped speckles with sufficient information for recovery can be captured by the detector.

By employing the data preprocessing and performing the de-mixing procedure, one can get nearly pure speckles for each separated target from the chaotic mixtures (see Fig. 4(c)). Different targets far beyond the OME range can be reconstructed efficiently by using ICA combined with iterative PR algorithm. In contrast, traditional PR method fails to reconstruct the object (see Fig. 4(d)), whereas, the proposed de-mixing model shows quite well clear results (see Fig. 4(e)).

To verify the proposed strategy for imaging beyond 3D OME, the home-designed object (see Fig. 3 (b), Fig. 5 (c)) consists of three targets printed on a film separately localized at different depths and positions using 3D printing framework are used. The equal central distance among targets in $x$-$y$ direction is 17mm and the depth separation in $z$ direction is 10mm. The OME range dictates the axial range which can be described as [23]:

$$\Delta Z_{FOV} \approx 2\Delta X_{FOV} \cdot R / D = k \Delta X_{FOV}, \qquad (8)$$

where $R$ is the distance between the ground glass and the point source, and $D$ denotes the 'scattering lens' aperture.

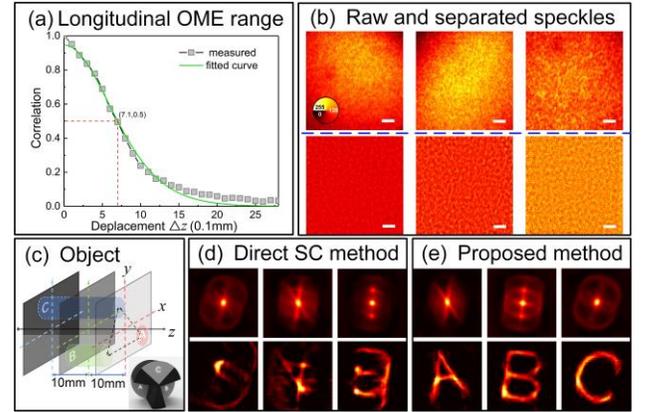

Fig. 5. Experiment results of multi-targets imaging beyond 3D OME range through the scattering layer. Scale bars are 2mm.

In the experiment, the point source used for the longitudinal OME measurement are shifted along $z$ direction with a fixed step 0.1mm considering the scaling factor correction [24]. As shown in Fig. 5(a), the HWHM for the correlation curve is 0.71mm, thereby, the effective $\Delta Z_{FOV}$ range is 1.42mm. Actually, the axial FOV is 7 times of the conventional one and six-fold of the lateral OME range can be achieved, simultaneously. Raw captured images and the de-mixed speckles separated using a blue dashed line are shown in Fig. 5(b). Fig. 5(d) and 5(e) show the ACs from 5(b) using chaotic and de-mixed speckle patterns, respectively, along with their corresponding reconstruction results. The classic SC method fails to resolve any discernable image of one of the targets directly, let along the whole view targets localized at different spatial positions.

For image restoration, we quantitatively define the degree of successful speckle separation as the cross-correlation between the source and separated speckle signals:

$$CC = \frac{\sigma_{xy}}{\sigma_x \sigma_y} = \frac{E\left[(x-\mu_x)(y-\mu_y)\right]}{\sqrt{E\left[(x-\mu_x)\right]^2 E\left[(y-\mu_y)\right]^2}}, \quad (9)$$

where *x, y* denote total intensities of two speckle vectors of the source and separated signals, $\mu_{x,y}$ are their mean values, $\sigma_{x,y}$ are their standard deviations, and $E[\cdot]$ is the expectation operator. The best speckle separation is obtained when $CC = 1$.

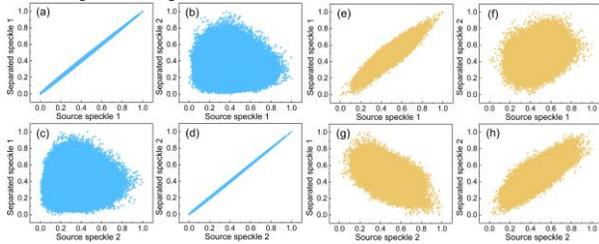

Fig. 6. Correlation plots between one of the separated speckle signals and one of the source signals for simulation results (a-d) and experimental results (e-h).

Figure 6 shows the correlation plots between one of the raw speckle signals and one of the separated signals in simulation and experiment. In each scatter diagrams, $10^6$ points are used. The corresponding pairs of the source and the separated signals show quite well correlation, i.e., the cross correlations of 0.9990 and 0.9993 for Figs. 6(a) and (d), respectively. However, the cross correlation of 0.1495 and 0.2099 for Figs. 6(c) and (d) are calculated. In experiment, the captured speckles usually suffer from thermal noise, which will lead to a decline in correlation. The observed correlation between the ground truth speckle and separated one is 0.8450 and 0.8265 in Figs. 6(e) and (h), respectively. However, in Figs. 6(f) and (g) the correlation coefficients are 0.3975, 0.3873. Therefore, we succeed blind signal separation of mixed speckle signals by ICA.

In conclusion, we proposed and demonstrated a scattering imaging technique beyond 3D OME using spatial decorrelation strategy and technique, for the first time, to the best of our knowledge, in which full FOV imaging of multi-targets at different depths were achieved by employing modulated illumination, ICA method and PR reconstruction. We experimentally demonstrated our method with four well-separated targets beyond angular OME range and three well-separated targets beyond 3D OME region without using guide stars or pre-calibrated PSFs. The proposed method was qualitatively compared with the traditional direct PR strategy numerically and experimentally. Moreover, the de-aliasing ability of ICA for separating mixed speckles on imaging efficiency was investigated by correlation criterions. The method is non-scanning and lens-less, and remarkably allows large FOV imaging through scattering and imaging of complex multi-targets (as long as every single target's size is within OME) at different spatial locations. This technique exhibits great potential in lens-less imaging, biomedical imaging, and no-reference 3D imaging. Further investigations on the capability of the proposed method with fewer measurements for extended object and multispectral imaging will be explored in the future.

**Funding.**

This work was financially supported by the National Natural Science Foundation of China (NSFC) (61975254) and the '111 project' (B17035).

**Acknowledgment**.
We would like to thank Shengcun Ma for the 3D-printing and Mingrui Xia for the helpful discussion.

**Contributions**
Project planning and idea proposed by: Jietao Liu, and Wei Li. Sample preparation, experiment measurements: Wei Li, and Jietao Liu. Manuscript written and figures prepared by Jietao Liu and Wei Li. All authors contributed to overall data analysis and scientific discussions.

**Appendix A**

In this appendix, we show here the recovery flow chart of imaging exceeding OME approach, where we feed in raw experiment speckles and the recovered objects either exceed lateral or 3D OME range come out. The recovery process is shown in Fig. 7.

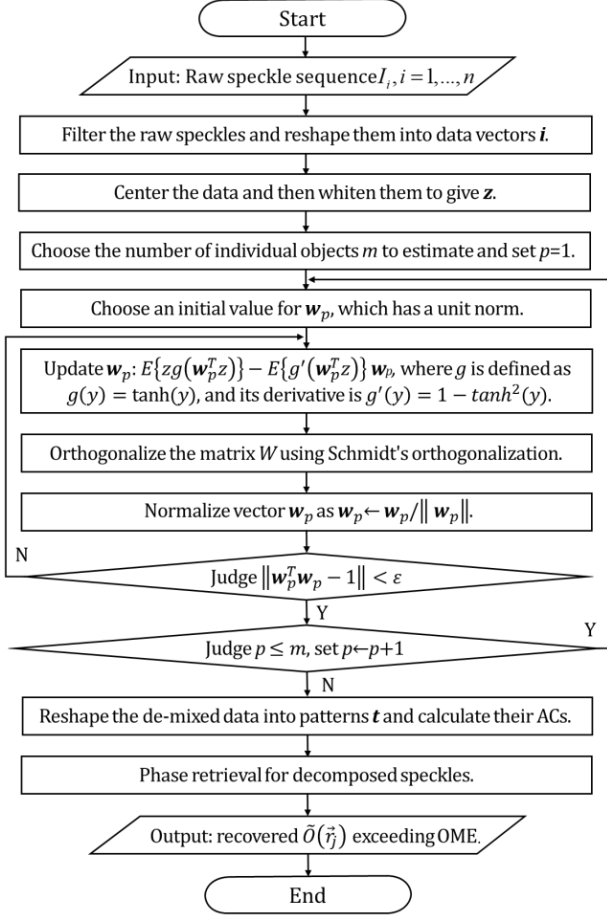

Fig. 7. Recovery processing flow chart of the proposed method for imaging exceeding OME range.